\documentclass[copyright,creativecommons]{eptcs}
\providecommand{\event}{HAS 2013} 
\usepackage{breakurl}             
\usepackage{graphicx}
\usepackage{amsfonts}
\usepackage{amsmath}
\usepackage{color}
\usepackage{amssymb}
\usepackage{amsthm}
\usepackage{amsbsy}
\usepackage{float}

\usepackage[usenames,dvipsnames]{xcolor}

\newtheorem{hyp}{Hypothesis}
\newtheorem{definition}{Definition}
\title{An Integrated Framework for Diagnosis and Prognosis of Hybrid Systems}
\author{Elodie Chanthery$^{1,2}$ \quad \quad \qquad Pauline Ribot$^{1,3}$ 
\institute{$^1$CNRS, LAAS, 7 avenue du colonel Roche, F-31400 Toulouse, France}
\institute{$^2$Univ de Toulouse, INSA, LAAS, F-31400 Toulouse, France}
\institute{$^3$Univ de Toulouse, UPS, LAAS, F-31400 Toulouse, France}
\email{elodie.chanthery@laas.fr, pauline.ribot@laas.fr}
}
\def\titlerunning{An Integrated Framework for Diagnosis and Prognosis of Hybrid Systems}
\def\authorrunning{Elodie Chanthery and Pauline Ribot}
\begin{document}
\maketitle

\begin{abstract}
Complex systems are naturally hybrid: their dynamic behavior is both continuous and discrete. For these systems, maintenance and repair are an increasing part of the total cost of final product. Efficient diagnosis and prognosis techniques have to be adopted to detect, isolate and anticipate faults. This paper presents an original integrated theoretical framework for diagnosis and prognosis of hybrid systems. The formalism used for hybrid diagnosis is enriched in order to be able to follow the evolution of an aging law for each fault of the system. The paper presents a methodology for interleaving diagnosis and prognosis in a hybrid framework. 
\end{abstract}

\section{Introduction}
The development of recent systems exhibits an increasing complexity of behaviors. This implies two problems. Firstly, it is becoming increasingly difficult to ignore the fact that most systems are hybrid: they present both continuous and discrete behaviors. Secondly, systems are exposed to failures. Maintenance and repair became an important part of the total cost of final product. Efficient diagnosis and prognosis techniques have to be adopted to detect, isolate and anticipate faults. 

Health monitoring of industrial systems aims at determining the health state of systems at any time in order to optimize their functioning and act in case of malfunctions. Diagnosis helps to determine the current health state of a system. Malfunctions or failures may be anticipated by a prognostic reasoning on the system. No definition of prognosis is really stated in the scientific community. It is more often related to the health state prediction in the future or to the calculation of the remaining useful life (RUL). This temporal prediction gives the date at which the system is not operational anymore and must to be repaired. When the system is in such a state, it is said to be in a \emph{failure mode}. Before this state, it can be either in a \emph{nominal mode}, or after the occurrence of one or several faults in a \emph{faulty mode} or degraded mode. Prognosis requires the knowledge of the current health state of the system through a diagnosis, hence the need of a health monitoring method 
integrating diagnosis and prognosis. 

This paper proposes to use the techniques of model-based diagnosis and prognosis in the framework of hybrid systems.  We propose to enrich the commonly used modeling framework for hybrid systems with available knowledge about aging or degradation of the system. Systems are continuously degrading according to operational conditions. 
According to information available on the system, it is possible to establish laws of physical aging or time-dependent probabilities of fault based on the feedback. This temporal and/or stochastic information should be taken into account in the model of the hybrid system.

This paper first gives a brief overview of the related work in diagnosis and prognosis on hybrid systems. Section 3 describes the formal background about the formalism used for modeling hybrid systems and for aging modeling. After the introduction of a new modeling framework, Section 4 presents a methodology for interleaved diagnosis and prognosis on hybrid systems. Section 6 concludes the paper and proposes some future work. 

\section{Related Work}

There has been considerable work on diagnosis of hybrid systems on one hand, and on prognosis on the other hand. However, to the best of our knowledge, very few studies succeed in coupling diagnosis and prognosis and the authors could not find related work dealing with prognosis on hybrid systems.  

A formal generic modeling framework for a complex system is presented in~\cite{Ribot09b} that encapsulates the knowledge used by diagnosis and prognosis. In this work, the authors establish a coupling of diagnosis and prognosis based on a characterization of complex system modes but no algorithm and implementation have been proposed. Another approach has been proposed in~\cite{Roychoudhury11}. The authors proposes a common framework for diagnosis and prognosis thanks to a state representation that describes the nominal behavior of the system and fault progression. However, there is absolutely no hybrid or discrete aspect in this work. The model used is a state model that specifies the system behavior in nominal model and in faulty modes. A parameter vector and an associated evolution equation allow to represent fault progression over time. The method consists in building an observer from the nominal behavior to perform fault detection. The identification is made from a set of observers that are built for each 
fault. Prognosis consists in predicting the remaining useful life (RUL) for each fault using an estimator based on a fault progression model. 

Of all related work on discrete events systems, we have not found even a handful of them that deals with prognosis development. Most of the works on discrete event systems consider prognosis as a prediction of an event trajectory~(\cite{Cao89}) or fault event occurrences~(\cite{Genc06a}). The term "predictability" of a fault event introduced by~\cite{Cao89} is based on the system observability property. It is clearly related to the diagnosability notion in discrete event systems: "it is certain that a critical event will take place". \cite{Genc09} demonstrates that any predictable event is diagnosable. An extension for the prediction of event patterns is proposed in~\cite{Jeron08}. In these studies, the system model is a classical automaton in which only ordered, undated and without delay event sequences are considered. 

To perform prognosis, it is required to take the temporal aspect into account to compute the RUL of the system. Only~\cite{Khoumsi09} uses a timed automata (TA) in order to prognose a fault event on the system. Clock ticks are added to transitions of the TA to determine the dated trajectories leading to fault events. No notion of uncertainty (neither by mean of probabilities nor intervals) is taken into account in these timed automata. However, uncertainty is intrinsically linked to prognosis. 

In~\cite{Zemouri06}, the evolution of the system operating state is modeled by a stochastic timed automaton (STA). A stochastic distribution $f(t)$ is associated to each transition of the automaton. The distribution $f(t)$ gives the probability of occurrence for $x_{j+1}$ at time $t_{j+1}$ after the occurrence of $x_{j}$ at time $t_{j}$. In this study, events occurring in the system are represented in the states of the stochastic timed automaton that does not allow to take the hybrid dynamics of the system into account. 

\cite{Castaneda10} proposes a stochastic hybrid automaton to evaluate the system dynamic reliability. 
The stochastic hybrid automaton represents the possible behavioral modes of the system. The stochastic part helps to take faults and uncertainties about system knowledge into account. The system switches from one mode to another with events that may be deterministic or stochastic. Stochastic events occur when a threshold on their probability law has been reached. In this study, stochastic transitions have a constant rate. 
The model is simulated to obtain availability and reliability defined as the probabilistic evaluation of the hybrid system failure.



\section{Background}

This section recalls formal background about the formalism used for modeling hybrid systems and also for aging modeling.

\subsection{Hybrid formalism}

The modeling framework that is adopted for hybrid systems is based on a hybrid automaton (\cite{Hen1996}). The hybrid automaton is defined as a tuple $S=(\zeta,Q,\Sigma,T,C,(q_0,\zeta_0))$
where:
\begin{itemize}
 \item $\zeta$ is the set of continuous variables that comprises input variables $u(t) \in R^{n_u}$, state variables $x(t) \in R^{n_x}$, and output variables $y(t) \in R^{n_y}$. The set of directly measured variables is denoted by $\zeta_{OBS}$.
 \item $Q$ is the set of discrete system states. Each state $q_i \in Q $ represents a behavioral mode of the system. It includes nominal and 
anticipated faulty modes, including failure modes. The anticipated faulty modes are faulty modes that are known to be possible on the system. The unknown mode can be added to model all the non-anticipated faulty situations. 
 \item $\Sigma$ is the set of events that correspond to discrete control inputs, autonomous mode changes and fault occurrences. $\Sigma
=\Sigma_{uo} \cup \Sigma_o $, where $\Sigma_{o} \subseteq \Sigma$ is the set of observable events and $\Sigma_{uo} \subseteq \Sigma$ 
is the set of unobservable events. 
 \item $T \subseteq Q \times \Sigma \rightarrow Q $ is the partial transition function. The transition from mode $q_i$ 
to mode $q_j$ with associated event $\sigma_{ij}$ is noted $t(q_i,\sigma_{ij},q_j)$ and we have $T(q_i,\sigma_{ij})=q_j$.
 $T$ also denotes the set of transitions.
 \item $C=\bigcup_{i} C_i$ is the set of system constraints linking continuous variables. $C_i$ denotes the set of constraints 
associated to the mode $q_i$. $C$ represents the set of differential and algebraic equations modeling the continuous behavior 
of the system. The continuous behavior in each mode is assumed to be linear.
 \item $(\zeta_0,q_0)\in \zeta \times Q $, is the initial condition.\\
\end{itemize}

\begin{figure}[h]
 \centering
 \includegraphics[width=7cm]{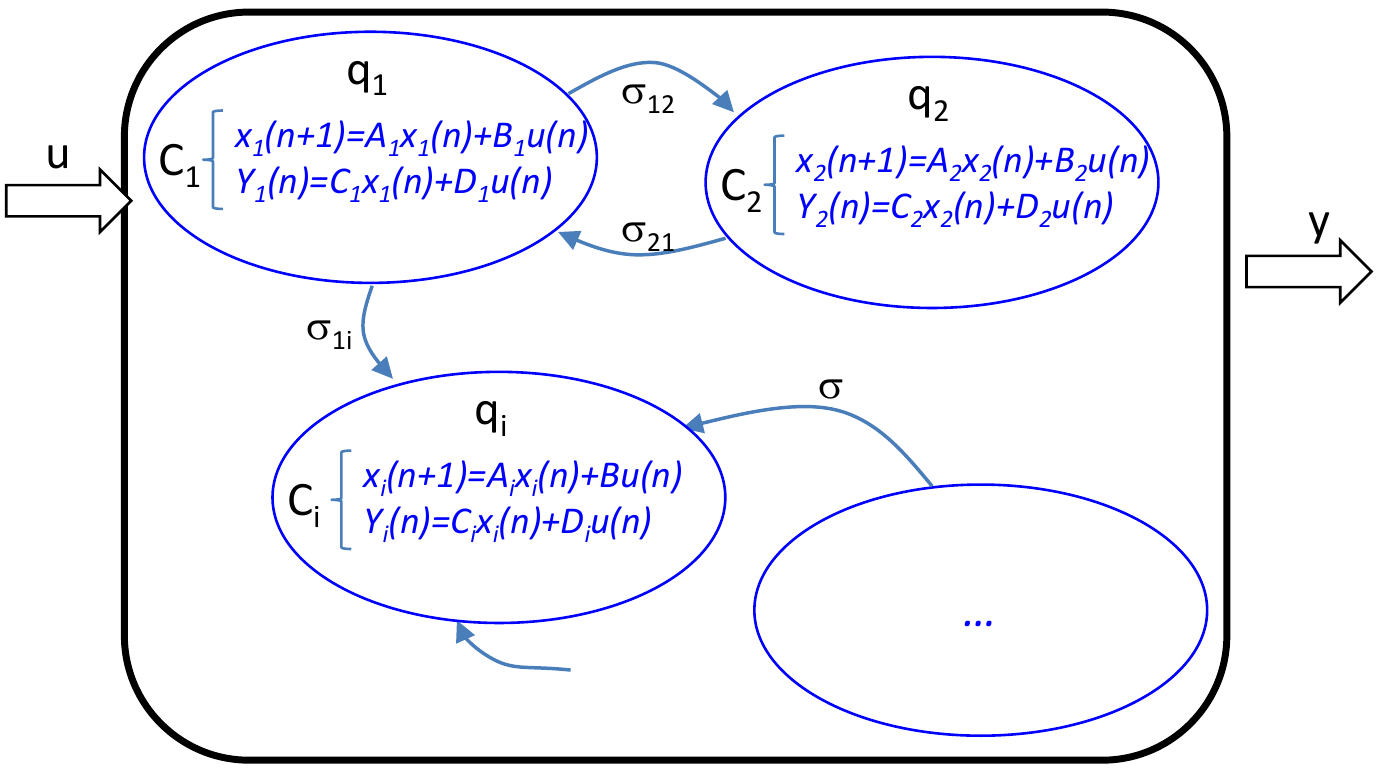}
 \caption{Example of an hybrid system}
 \label{fig:hybridsystem}
\end{figure}

The occurrence of a fault is modeled by a discrete event $f_i\in \Sigma_{F}$. $\Sigma_{F}$ is the set of fault events associated to the anticipated faults of $F$. Without loss of generality it is assumed that $\Sigma_F \subseteq \Sigma_{uo}$. The discrete part of  the hybrid automaton is given by $M=(Q,\Sigma,T,q_0)$, which is called the \emph{underlying discrete event system (DES)} and the continuous behavior of the hybrid system is modeled by the so-called \emph{underlying multimode system} $\Xi=(\zeta,Q,C,\zeta_0)$. An example of a hybrid system is given in Figure~\ref{fig:hybridsystem}. 

This hybrid automaton describes the set of knowledge useful to achieve model-based diagnosis. In order to perform prognosis, it is necessary to enrich the hybrid model by adding the available knowledge about the aging or the degradation of the system components. The next section presents a way to take the uncertainty on the degradation function into account by introducing probability measures for each state that represents a mode of the system. 

\subsection{Aging modeling}
\label{subsec:aging}

The modeling framework that is adopted for the system degradation is based on the Weibull probabilistic model~(\cite{Ribot11}). A particular way for representing the remaining useful life of systems is to establish a fault probability from reliability analyses at different stress levels (operating conditions)~(\cite{Hall03,Vachtsevanos06}). Stress is defined as the set of internal and external conditions/factors that may have an impact on the system behavior. The parametrized Weibull model is often used in reliability for life data analyses due to its flexibility~(\cite{Ferreiro08}):
\begin{equation}
W(t, \beta,\eta,\gamma)=  \frac{\beta}{\eta} \Big( \frac{t - \gamma}{\eta} \Big) ^{(\beta - 1)} e^{-( \frac{t -\gamma}{\eta})^\beta}
\end{equation}
where $t \geq 0$, $\beta \geq 0$, $\eta \geq 0 $ and $\gamma \in [-\infty;\infty]$. The scale characteristic $\eta$ defines the characteristic life of the system and corresponds to the mean life expectancy for a studied population sample. The shape characteristic $\beta$ modifies the probability density function (pdf) nature and allows to model the different life phases of a system defined by the idealized bathtub curve of reliability. The location characteristic $\gamma$ shifts the curve from the origin. It defines the system minimal life. The case $\gamma>0$ means the fault probability is zero until a date $\gamma$. In most cases, we assume $\gamma=0$. The characteristic $\eta$ is stress-dependent while $\beta$ is assumed to remain constant across different stress levels.

Weibull characteristics model the aging evolution of a system that lead to a fault $f_j$ in a behavioral mode $q_i$ and totally define the fault probability distribution $f^{q_i}_j$:
\begin{equation}
f^{q_i}_j(t)= \int_{0}^{t} W(t, \beta^{q_i}_j,\eta^{q_i}_j,\gamma^{q_i}_j)dt.
\end{equation}
The fault probability density function $W(t, \beta^{q_i}_j, \eta^{q_i}_j, \gamma^{q_i}_j)$ has to give at any time the probability that the fault $f_j$ occurs in the system from a mode $q_i$. Weibull characteristics $\beta^{q_i}_j$ and $\eta^{q_i}_j$ are fixed by the mode $q_i$ of the system. The location characteristic $\gamma^{q_i}_j$ can be used to memorize the degradation evolution of the system in the past modes from the operation start of the system~(\cite{Ribot11}). At first, the system is in a nominal mode ${q_0}$ and we suppose that $\forall f_j, ~~\gamma^{q_0}_j=0$ as previously explained. This characteristic $\gamma^{q_i}_j$ will be modified to take degradation in each behavioral mode into account during the system operation. 

The occurrence date $d_{f_j}$ of a fault event $f_j$ for the system in mode $q_i$ can be determined from a decision criterion $P_{max}$ that corresponds to a probability threshold beyond which the risk becomes unacceptable: 
\begin{equation}
d_{f_j}\mbox{ such that }\int_{0}^{d_{f_j}} W(t,\beta^{q_i}_j,\eta^{q_i}_j,\gamma^{q_i}_j)dt=P_{max}.
\end{equation}

\section{Methodology for interleaved diagnosis and prognosis}

This section presents the new framework we propose to integrate both diagnosis and prognosis processes on hybrid systems. The goal of this work is to integrate some information into the hybrid model in order to be able to perform both diagnosis and prognosis. To this end, we adopt the formalism of hybrid automata and enrich this formalism with data useful for the prognosis. After an overview of the methodology, some important hypotheses are given. We then describe how diagnosis and prognosis can be performed and how they are interleaved.   

\subsection{Methodology overview}

\subsubsection{The enriched modeling framework}

The hybrid system is described as an enriched hybrid automaton $S^+=(\zeta,Q,\Sigma,T,C,\mathcal{F}, (q_0,\zeta_0))$. 
In each mode $q_i$, the system is subject to different aging laws. 
The set of aging laws is supposed to be accurately known. 
$\mathcal{F}=\{ F^{q_i}, i\in\{1,\ldots,card(Q)\}\}$ is the set of aging laws associated to behavioral modes or the system. ${F}^{q_i}$ is a vector of aging laws for each anticipated fault in the mode $q_i$. For example, in a system where $N_F$ faults are considered: 
\begin{equation}
 F^{q_i}(t) = \left[ \begin{array}{c}
               f^{q_i}_1(t)\\
f^{q_i}_2(t)\\
\ldots \\
f^{q_i}_{N_F}(t)
              \end{array} \right]
\end{equation}

where $f^{q_i}_j$ represents the probability distribution of the fault $f_j$ at any time in mode $q_i$. 

It can be noticed that as opposed to \cite{Ribot11}, the hybrid automaton we propose represents behavioral modes and not operational modes based on function availability.

\subsubsection{Methodology steps}
The diagnosis method for hybrid systems that is adopted for our approach is the one developed in (\cite{Bay2009},  \cite{BaTrOlSp08}).  It interlinks a standard diagnosis method for continuous systems, namely the parity space method, and a standard diagnosis method for DES, namely the diagnoser method (\cite{Sam:95}). 
The hybrid diagnoser takes as input the hybrid automaton part $S$ of $S^+$. It needs the set of observations on the system on-line. 
The diagnosis part of the methodology may be decomposed into three parts: 
\begin{itemize}
 \item diagnose the continuous part of the system,
 \item abstract the continuous part in terms of discrete events and enrich the discrete part of the system with discrete events that come from the abstraction of the continuous part,
 \item then apply the diagnoser method on the resulting discrete event system. 
\end{itemize}

Prognosis takes as input the information of aging laws in $S^+$. 
On-line, the prognosis process is updating the aging laws for the system according to the operation time in each behavioral mode. After each observable event, the appropriate aging laws are selected according to  the mode that is estimated from diagnosis and the fault probability value reached in previous modes. 
The goal of the prognosis is to compute the most likely sequence of dated faulty modes leading to the system failure.

\subsection{Hypotheses and definitions}

Our main contribution is to consider that diagnosis and prognosis are no longer two processes viewed separately. The goal is that they interact and enrich each other. It is first necessary to explicit some important hypotheses in our work.

First it is supposed that the system is diagnosed after the occurrence of each observable event. As diagnosis consists in monitoring the diagnoser, the diagnosis computation duration can be considered as instantaneous. It is also supposed that between two observable events, both diagnosis and prognosis can be performed. 
Let $t_{k}$ be an occurrence of an observable event and $t_{{k+1}}$ the next one. Let $CT_p$ be the computation time for prognosis.

\begin{hyp}
The computation time of the prognosis process is smaller than the interval between two occurrences of observable events.
\begin{equation}
 CT_p \leq (t_{{k+1}} - t_{k})
\end{equation}
\end{hyp}

The second hypothesis is on aging laws associated to each anticipated fault of the system. 
\begin{hyp}
 The aging law of each component is supposed to be continuous over time. 
\end{hyp}

The consequence of this hypothesis is that the initial condition for an aging law at time $t_{{k+1}}^+$ is the value at $t_{{k+1}}^-$, when the system has not yet commuted between two modes.

\begin{figure}[h]
 \centering
 \includegraphics[width=9cm]{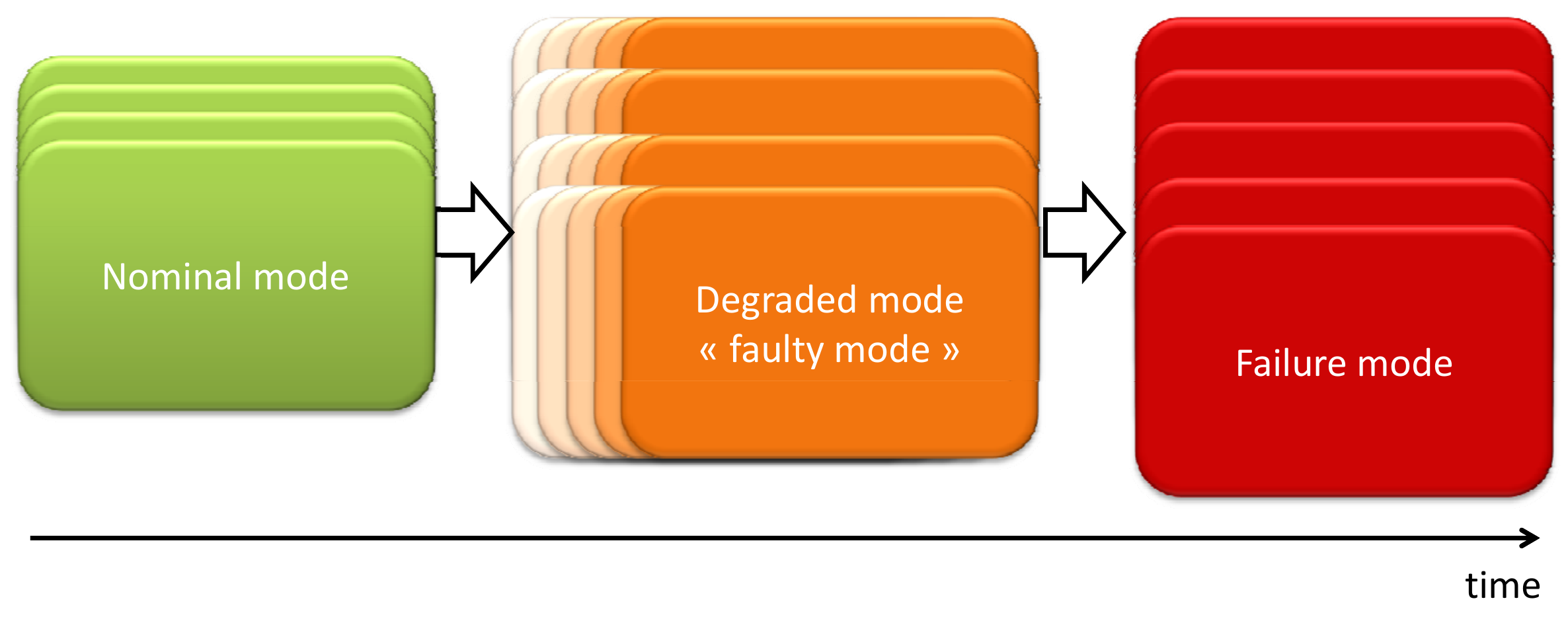}
 \label{fig:Evolution}
\caption{Unidirectional system evolution without maintenance or repair action}
\end{figure}

Finally, it is supposed in this work that starting from a nominal behavior, a fault may occur at any time. Therefore, in our models each fault can occur from each nominal state. 
Moreover, faults are supposed to be permanent: it means that once a fault has occurred, the system evolves in what is called a \emph{faulty mode}. This degradation can evolve into a worst degraded mode. 
Finally, when the system is not operational anymore, it is said to be in a \emph{failure mode}. Without maintenance or repair action, the evolution of a system is then supposed to be unidirectional. 
This system evolution is illustrated in Figure~\ref{fig:Evolution}.

\begin{figure}[h]
 \centering
 \includegraphics[width=9cm]{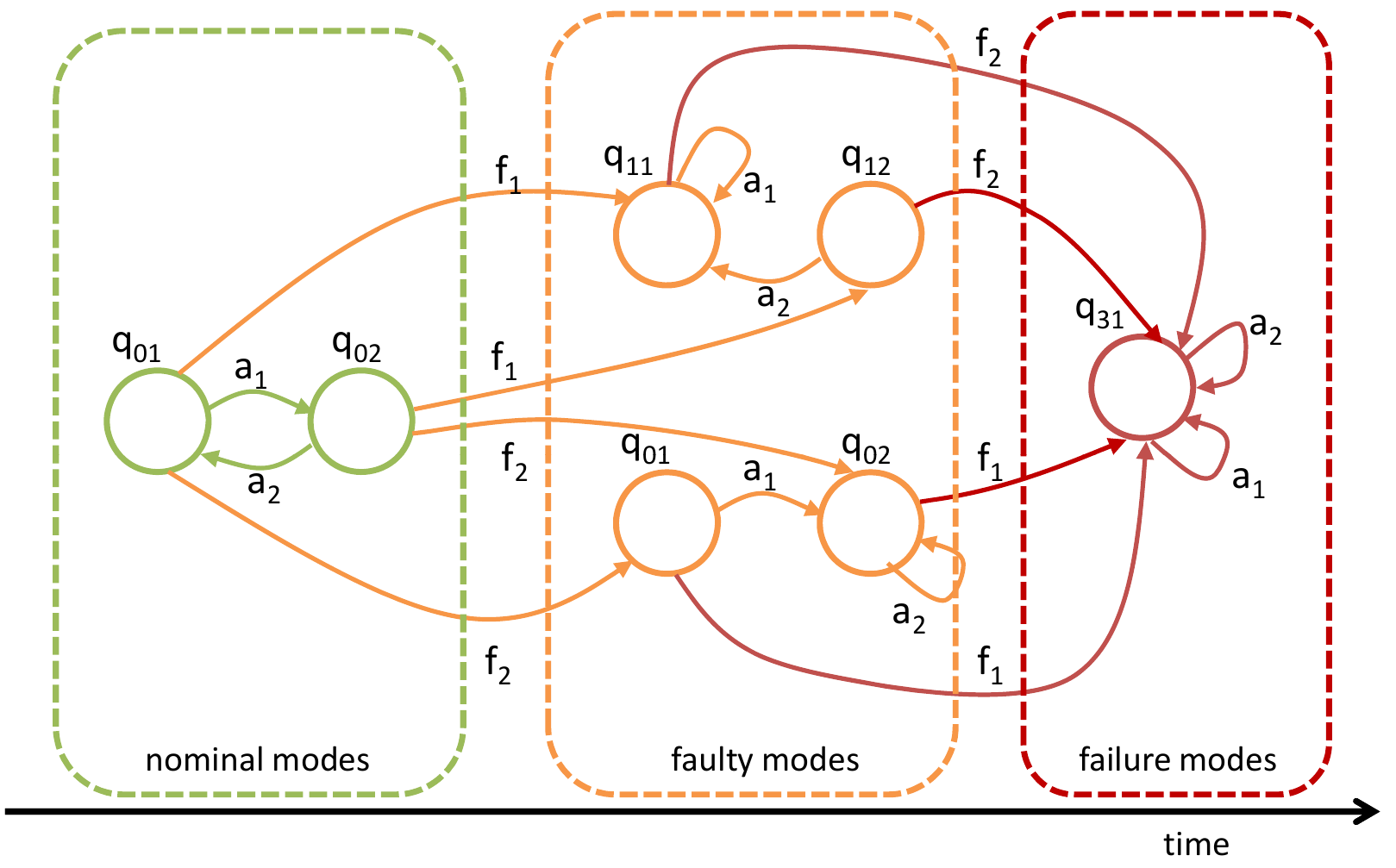}
 \caption{Example for a system with 2 nominal modes}
 \label{fig:Evolutionex}
\end{figure}

For example, for a system with two nominal modes $q_{01}$, $q_{02}$, two possible actions $a_1$, $a_2$ that are observable events, and two faults $f_1$ and $f_2$, a possible model is given in Figure~\ref{fig:Evolutionex}. This system is in a failure mode when $f_1$ and $f_2$ have occurred. If only one fault occurred, then the system is in a faulty mode. The combination of faults leading to a failure can be established from a fault tree analysis (\cite{Rausand04}). With this analysis and the sequence of fault dates predicted by prognosis, it is simple to obtain the system RUL that corresponds to the remaining time until the system failure. This fault analysis allows to link our prognosis definition to the one commonly used in the PHM community (Prognostics and Health Management). 

The next sections describe in detail how diagnosis and prognosis are developed in our methodology.  

\subsection{Diagnosis of the continuous part of the system}

Consistency tests take the form of a set of analytical redundancy relations (ARR). For the linear case, it is possible to computes ARRs by using the parity space approach (\cite{Sta2001}). 
In our case, an ARR set is associated to each mode $q_i$ and is denoted $ARR_i$. An
$ARR_{ij}$ can be expressed as $r_{ij}=0$, where $r_{ij}$ is called the residual of the ARR. Since ARRs are constraints that only contain
observable variables, they can be evaluated on-line with the incoming observations given by the sensors, allowing to check the consistency of the
observed system behavior with the predicted one. ARRs are satisfied if the observations satisfy the model constraints, in which case the associated
residuals are zero. In the opposite case, all or some residuals may be non-zero. The set of residuals in mode $q_i$ hence result in a local boolean 
fault indicator tuple.
\begin{equation} r_{ij} =
\left \{
\begin{array}{rl}
0 & \mbox{when } ARR_{ij} \mbox{ is satisfied} \\
1 & \mbox{otherwise}
\end{array}\right.
\end{equation}
$j=1, ..., N_{ARR(q_i)}$, where $N_{ARR}(q_{i})$ is the number of ARRs/residuals associated to mode $q_i$.\\ 

Due to the fact that there are different modes, we need to define the vector of residuals of mode $q_k$ computed with the observations $\zeta_{OBS}^j$ obtained when the system is in mode $q_j$. This vector is called the $q_k$-mirror signature of mode $q_j$. It is the signature of mode $q_j$ seen in mode $q_k$.

\begin{definition}[Mirror Signature]
Given the vector $R_k=[r_{k1}, r_{k2}, ..., r_{kN_{ARR(q_k)}} ]^T$ of system residuals in mode $q_k$, the $q_k$-mirror signature of mode $q_j$ is given by the vector $S_{j/k}=[s_{1_{j/k}}, ..., s_{N_{ARR(q_k)_{j/k}}}]^T=[R_k(\zeta_{OBS}^j)]^T$. 
\end{definition} 

\begin{definition}[Mode Signature]
The mode signature of a mode $q_j$ is the vector obtained by the concatenation of all its mirror signatures, $Sig(q_j)=[S^{T}_{j/1}, S^{T}_{j/2}, ..., S^{T}_{j/j}, ..., S^{T}_{j/m}]^T.$ 
\end{definition}

Since residuals have been designed for every mode separately and that there exists noise, this may cause false alarms. For solving this problem, the diagnoser has to use an on-line residual filter that takes as input the residual values and generates clean boolean indicators as output. Two modes $q_i$ and $q_j$, $(q_i,q_j) \in Q^2$ are diagnosable if their continuous dynamics defined by the state-space matrices $A_i$, $B_i$, $C_i$, $D_i$ and $A_j$, $B_j$, $C_j$, $D_j$ are distinct: their cleaned residuals are not identical i.e. their mode signatures are different. If two modes are not diagnosable, they are merged into the same group of modes.

\subsection{Event-based abstraction of the continuous dynamics and diagnoser's build}
The idea of the hybrid system diagnosis is to capture both the continuous dynamics and the discrete dynamics within the same mathematical object. For this purpose, the original automaton of the hybrid system $M=(Q,\Sigma,T,q_0)$ is enriched with specific observable events that are generated from the mode signatures. A specific transition labeled with a specific \emph{signature-event} is introduced between two modes when they have different mode signatures. The change of signature triggers the occurrence of the associated signature-event. The resulting automaton is called the \emph{Behavior Automaton} (BA) of the hybrid system.

Diagnosis is performed thanks to a specific finite machine called a \emph{diagnoser}. The diagnoser is built from the BA following the approach described in (\cite{Sam:95}). The task of building such diagnoser is not easy because it requires to browse the entire graph representing the BA automaton. To this end, the tool DiaDES\footnote{http://www.laas.fr/$\sim$ypencole/DiaDes} from LAAS, Toulouse, can be used. It allows the automatical generation of the diagnoser.

At each time step, the hybrid diagnoser provides the possible diagnoses, i.e. the possible states of the BA and the fault event(s) that have occurred. To this end, it computes the filtered residuals, evaluates the observed mode signature, and generates the corresponding signature-event. Thanks to the signature-events and the other observable events that occur on the system, the corresponding branch of the diagnoser is explored on-line and the diagnosis results are delivered. The hybrid diagnoser provides possible diagnoses at each clock tick $t_k$: 
\begin{equation}
\Delta_k = \left[ \begin{array}{c}
\Delta_k^1 \\
\Delta_k^2 \\
\vdots \\
\Delta_k^D
\end{array} \right]
\end{equation}
where $D$ is the number of diagnosis hypotheses. In case of a non-diagnosable system, a lot of ambiguity remains in the diagnoser answer. For this purpose, interleaving diagnosis and prognosis is clearly useful.

\subsection{Prognosis}

In order to determine the RUL of the system, the remaining time until a fault occurs has to be evaluated from aging models $\mathcal{F}$. The prognostic process has to compute and update the fault pdf of the system according to the mode of the system (that corresponds to a stress condition) estimated from diagnosis.

\subsubsection{Prognosis initialization}

\paragraph{Selection of aging models}

The appropriate aging models of the system $\mathcal{F}^{q_i}$ are selected according to the system mode $q_i$ estimated from the current diagnosis $\Delta_k$. Initially, the system is assumed to be in nominal mode $q_0$, the vector $F^{q_0}$ that models the system aging according to each fault $f_i$ with Weibull pdf $W(t,\beta_i^{q_0},\eta_i^{q_0},\gamma_i^{q_0})$ is considered. The characteristic $\gamma_i^{q_0}$ is supposed to be null as explained in Section~\ref{subsec:aging}. The Weibull pdf allows us to obtain the probability of each fault $f_i$ at any time in initial mode $q_0$. 

\paragraph{Determination of the next possible fault}

The occurrence of fault event $f_i$ at time $d_{f_i}$ is predicted from mode $q_0$ using its probability function $f_i(t)$ and an unacceptable probability threshold $P_{max}$:  \begin{equation}
d_{f_i} \mbox{ such that  } \int_0^{d_{f_i}} W(t,\beta_i^{q_0},\eta_i^{q_0},\gamma_i^{q_0})dt=P_{max}.
\end{equation}
It must be noticed that $f_i$ stands for the fault event and $f^{q_0}_i(t)$ is a probability distribution associated to the occurrence of fault $f_i$ from mode $q_0$. 
The occurrence date $d_{f_i}$ is then computed for each anticipated fault $f_i$ from aging models in the mode $q_0$. The next possible fault $f_{min_1}$ is determined from the minimal predicted fault date that is noted $d_{min_1}$: $d_{min_1} = \min_{i=\{1,\ldots,N\}}(d_{f_i})$. 

\subsubsection{Fault propagation in predictions}

At $d_{min_1}$, the system is predicted to switch into fault mode $q_{f_{min_1}}$. New aging models in fault mode $q_{f_{min_1}}$ (described by the Weibull pdf $W(t,\beta_i^{q_{f_{min_1}}},\eta_i^{q_{f_{min_1}}},\gamma_i^{q_{f_{min_1}}})$) have to be considered for each fault $f_i$ whose occurrence date is superior to $d_{min_1}$. The mode change predicted at $d_{min_1}$ may result in a modification of fault dates $d_{f_i}$ that have been previously computed. The value of characteristic $\gamma_i^{q_{f_{min_1}}}$ of aging models in the future mode $q_{f_{min_1}}$ is computed from the fault probability $P_{f_i}^1$ the system could have reached for fault $f_i$ at predicted commutation time $d_{min_1}$:
\begin{equation}
P_{f_i}^1= \int_0^{d_{min_1}} W(t,\beta_i^{q_0}, \eta_i^{q_0}, \gamma_i^{q_0}) \mathrm{d}t,
\end{equation}
\begin{equation}
\gamma_i^{q_{f_{min_1}}} = (d_{min_1}-\delta) \mbox{ such that } \int_0^{\delta} W(t,\beta_i^{q_{f_{min_1}}},\eta_i^{q_{f_{min_1}}}, 0) \mathrm{d}t = P_{f_i}^1.
\end{equation}  
Characteristic $\gamma_i^{q_{f_{min_1}}}$ allows to memorize the system aging in past modes. 
The date $d_{f_i}$ of fault occurrences are modified as follows:
\begin{equation}
d_{f_i} \mbox{ such that } [ P_{f_i}^1+\int_{d_{min_1}}^{d_{f_i}} W(t,\beta_i^{q_{f_{min_1}}},\eta_i^{q_{f_{min_1}}},\gamma_i^{q_{f_{min_1}}})dt = P_{max}]. 
\end{equation}
The next possible fault $f_{min_2}$ is determined from the minimal predicted fault date that is noted $d_{min_2}$: $d_{min_2} = \min_{i=\{1,\ldots,N\}\backslash min_1}(d_{f_i})$. Fault propagation is studied as explained above to compute $\gamma_i^{q_{f_{min_2}}}$ for faults that have not reached their probability threshold at $d_{min_2}$ using new aging models for mode $q_{f_{min_2}}$. This procedure stops when prognosis has predicted a possible future sequence containing all possible anticipated faults.

To summarize, the prognosis process computes the most likely future sequence $\Pi_k^j$ of dated fault events according to a diagnosis hypothesis $\Delta_k^j$.  
\begin{equation}
\Pi_k^j = (\{f_{min_1}, t_{min_1}\}, \{f_{min_2}, t_{min_2}\}, \ldots, \{f_{min_N}, t_{min_N}\}).
\end{equation}
%

\subsubsection{Prognosis update with on-line diagnosis}
\begin{figure}[h]
 \centering
 \includegraphics[width=12cm]{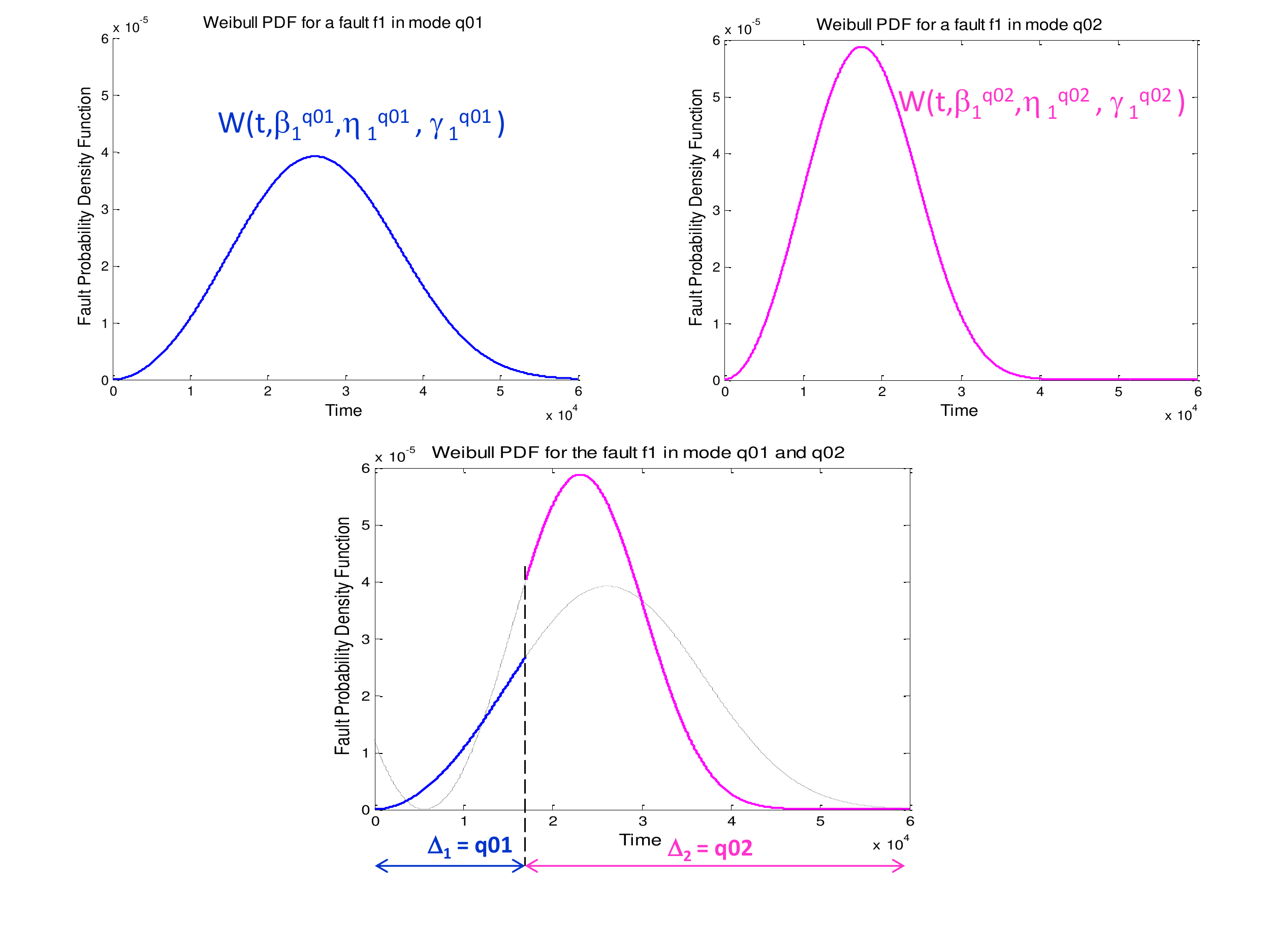}
\caption{ \label{fig:weibullpdf} Evolution of Weibull pdf for a fault $f_1$ according to the system behavioral mode}
\end{figure}

Every time a new observation is received at $t_k$, the diagnosis delivers new hypotheses on the behavioral mode of the system. Prognosis has to be performed considering aging models associated to each hypothesis $\Delta_k^j$ of the new belief state. Fault probabilities at $t_k$ are taken into account by computing new value for characteristic $\gamma_i^{\Delta_k^j}$ for aging models in mode $\Delta_k^j$.
For example, Figure \ref{fig:weibullpdf} illustrates how Weibull pdf associated to the fault event $f_1$ are modified to describe the system aging in two behavioral modes $q_{01}$ and $q_{02}$. 

\subsection{Interleaving diagnosis and prognosis processes}

To sum up, our methodology results in a new process that interleaves diagnosis and prognosis processes. This new process is called \emph{InterDP}.
\begin{figure}[h]
 \centering
 \includegraphics[width=12cm]{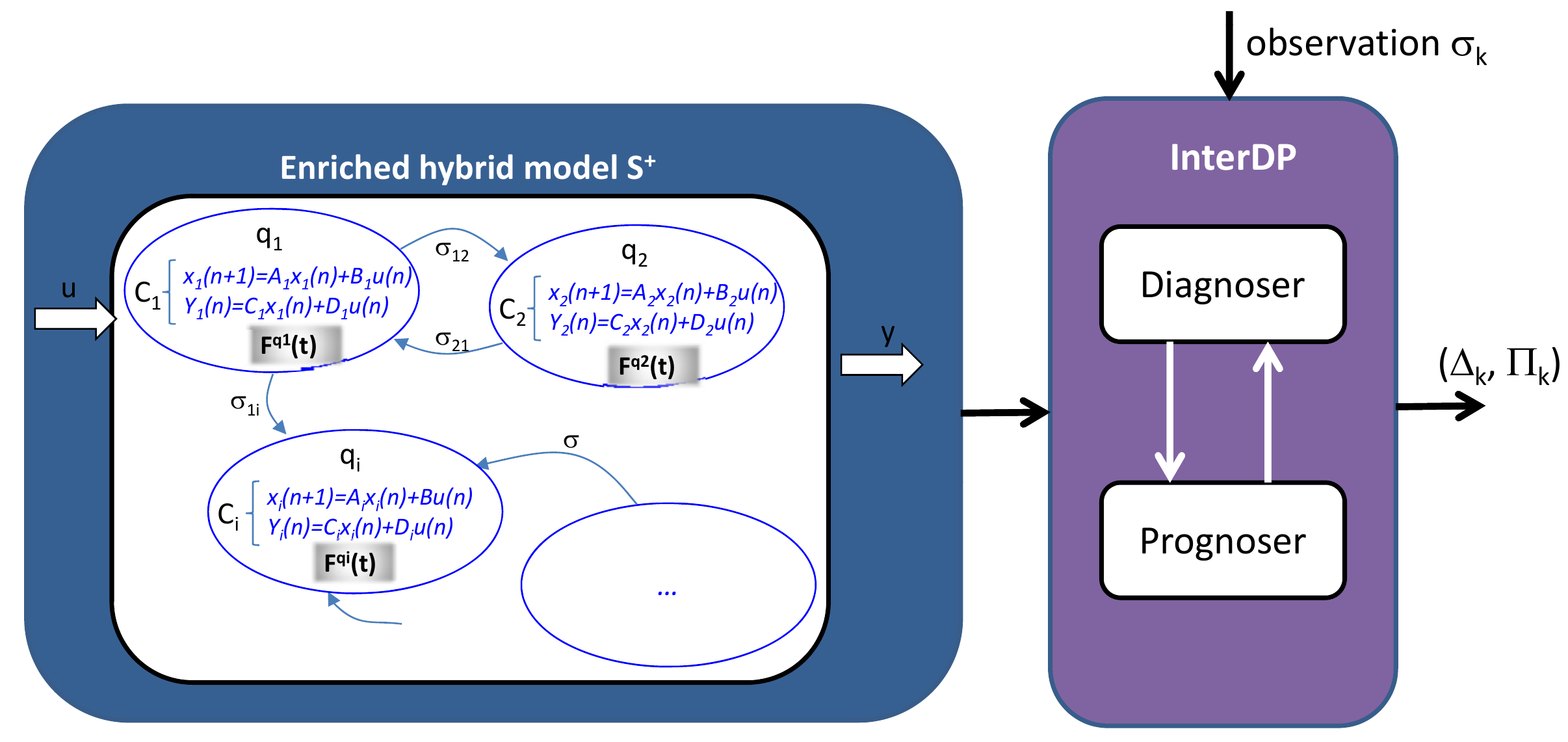}
\caption{\label{fig:InterDP}InterDP process interactions}
\end{figure}

The methodology is illustrated in Figure~\ref{fig:InterDP}. Inputs of the InterDP process are the enriched model defined for our methodology and observations. The output of InterDP at each clock tick $t_k$ is a couple~$(\Delta_k, \Pi_k)$. $\Delta_k$ is the output of the hybrid diagnoser. It is a diagnosis, i.e. a vector with the set of possible modes for the system: 
\begin{displaymath}
 \Delta_k = \left[ \begin{array}{c}
\Delta_k^1 \\
\Delta_k^2 \\
\vdots \\
\Delta_k^D
\end{array} \right]
\end{displaymath}
$\Delta_k$ can be seen as a belief state. If there are several possible modes, they may be sorted by decreasing probability for example. The probability of a mode may be given by the probabilities of the faults computed by the prognoser. $\Pi_k$ is the output of the prognoser:
\begin{displaymath}
 \Pi_k = \left[ \begin{array}{c}
\Pi_k^1 \\
\Pi_k^2 \\
\vdots \\
\Pi_k^D
\end{array} \right]
\end{displaymath}
For each $\Delta_k^i$, $\Pi_k^i$ represents the most likely sequence of dated faulty modes leading to the system failure: 
\begin{displaymath}
\Pi_k^j = (\{f_{min_1}, t_{min_1}\}, \ldots, \{f_{min_i}, t_{min_i}\}, \ldots, \{f_{min_N}, t_{min_N}\}).
\end{displaymath}
where $t_{min_i}$ is the date occurrence of fault $f_{min_i}$ and $N$ represents the number of degraded modes before the failure mode.



\section{Conclusion and future work}

This article presented an original theoretical framework  to follow the evolution of hybrid systems that encompasses both continuous and discrete event systems framework.
The paper proposed a new modeling framework for diagnosis and prognosis of hybrid systems by enriching the commonly used hybrid modeling with aging information. 
A methodology was introduced, that leads to the build of a new process named InterDP, that interleaves both diagnosis and prognosis processes. 
InterDP takes as input the enriched model of an hybrid system and provides as output a couple (diagnosis, prognosis) at each clock tick. 

In future work, we will implement this framework into a piece of
software that is used for the diagnosis of hybrid system for the
moment. The links between diagnosis and prognosis in case of ambiguity
will be further investigated. Methods that increase the amount of available
information concerning an ambiguous diagnosis are obviously
useful to focus on a reduce set of hypotheses~\cite{Rienmuller13}. 
Another idea is that diagnosis and prognosis processes feed each other: prognosis allows the diagnosis to be less ambiguous and thanks to the diagnosis, the prognosis process can refine its prediction.

\section{Bibliography}

\bibliographystyle{eptcs}
\bibliography{biblio,mabiblio}

\end{document}